\documentclass[12pt]{iopart}

\usepackage[latin1]{inputenc}
\usepackage{graphicx}
\usepackage{color}
\begin{document}

\title[Surface critical behaviour of the VISAW]{Surface critical behaviour of the vertex-interacting self-avoiding walk on the square lattice}

\author{D P  Foster and C Pinettes}

\address{Laboratoire de Physique Th\'eorique et Mod\'elisation
(CNRS UMR 8089), Universit\'e de Cergy-Pontoise, 2 ave A. Chauvin
95302 Cergy-Pontoise cedex, France}

\begin{abstract}
The phase diagram and surface critical behaviour of the vertex-interacting self-avoiding walk are examined using transfer matrix methods extended using DMRG and coupled with finite-size scaling. 
Particular attention is paid to the critical exponents at the ordinary and special points along the collapse transition line.  The question of the bulk exponents ($\nu$ and $\gamma$) is addressed, and the results found are at variance with  previously conjectured exact values.
\end{abstract}

\submitto{J Phys A}

\maketitle

\section{Introduction}

Polymers in dilute solution are known to undergo a collapse transition as the temperature or solvent quality is changed at what has come to be known as the $\Theta$-point\cite{flory}. Using universality arguments, it is reasonable to expect the thermodynamic behaviour of a lattice model to be the same as the continuum real system as long as the dimension of the system, basic symmetries and range of interactions are the same. Such lattice models (self-avoiding walks) have been used as models for real, linear polymers in solution for over three decades\cite{vanderbook}. 

The quality of the solvent may be introduced by the inclusion of short-ranged interactions in the model; typically an attractive energy is included for non-consecutive nearest-neighbour occupied lattice sites. This model is the standard Interacting Self-Avoiding Walk model (ISAW) or $\Theta$-point model\cite{flory, degennes75}. The model has been  shown to accurately predict the critical behaviour of a wide range of real linear polymers in solution, not only in the high-temperature phase, but also at the collapse transition, which occurs as  the temperature is lowered, at the $\Theta$ temperature.  The model is successful because it captures the strong entropic repulsion between different portions of the polymer chain (the self-avoidance), as well as the effect of the difference of affinity between monomer-monomer contacts and monomer-solvent contacts (attractive interaction). 

Whilst the relevant physical dimension in polymer physics would usually be $d=3$, the ISAW model has been much studied in two dimensions. This is partly motivated by the realisation that $d=3$ is the upper critical dimension of the collapse transition, and that the model in two dimensions provides an interesting playground.  In this paper we shall concentrate on the two-dimensional square lattice.

In the late eighties and early nineties there was much discussion about the universality class of the ISAW model, particularly with respect to the adsorption of the collapsing walk in the presence of an adsorbing wall\cite{ds,seno88,dsb, merovitchlima,dsc,vss91,foster92}. For a while there was an apparent contradiction between a slightly modified walk model on the hexagonal lattice (the $\Theta^\prime$ model) and the standard $\Theta$ model\cite{ds}. This contradiction arose in the surface exponents; the exact surface exponents from the $\Theta^\prime$ point model were not the same as those calculated numerically for the $\Theta$ model\cite{veal,ss88}.  The apparent 
contradiction was resolved when it was realised that the exact solution of the $\Theta^\prime$ model gives the exponents at the so called special point (where collapse and adsorption occur simultaneously) whilst the numerical calculations were at the ordinary point (where collapse occurs in the presence of the wall, but without adsorption)\cite{vss91}. This was verified for both models at the two different points using exact enumeration\cite{foster92}.

The debate over the $\Theta$ and $\Theta^\prime$ models opened the debate over to what extent the nature of the collapse transition depends on the details of the model. Different models were examined, and a range of collapse transitions were observed. Bl\"ote and Nienhuis introduced an $O(n)$ symmetric model which in the limit $n\to 0$ gives a bond self-avoiding walk model, which is allowed to visit sites more than once but not the lattice bonds\cite{blotenienuis}. The walk is not allowed to cross. Since the interactions are on the lattice vertex, we shall henceforth refer to this model as the vertex-interacting self-avoiding walk (VISAW). This model was shown to have a different collapse transition  than the 
$\Theta$ point model, with different bulk exponents\cite{wbn}; the correlation exponents $\nu=12/23$ for the VISAW compared to $4/7$ for the ISAW and the entropic exponent $\gamma=53/46$ compared to $\gamma=8/7$ for the ISAW. These exponents are conjectured based on a Bethe-Ansatz calculation of 
Izergin-Korepin model\cite{wbn}, and to the best of our knowledge have not been numerically tested since their conjecture. 

In recent years there has been a revival in another model with vertex interactions: the interacting 
self-avoiding trails (ISAT) model\cite{massih75}. 
This model corresponds to the VISAW in which the no-crossing constraint is relaxed. 
Evidence was presented by one of us that the $\nu$ exponent was also given by $\nu=12/23$, whilst $\gamma=22/23$\cite{F09}. A similar situation occurs in the ISAW on the Manhattan lattice, 
where the walk can only go one way down a row or column of lattice bonds, but the allowed direction alternates from one row (column) to the next. Here too the correlation length exponent is the same as the normal ISAW one, but $\gamma=6/7$ rather than $8/7$\cite{bradley89,bradley90}. 

Recently the surface exponents of the ISAT model were calculated using transfer matrix calculations\cite{F10}. We propose here to similarly calculate the surface critical behaviour of the VISAW model. 
In the case of the VISAW model, the no-crosssing constraint allows us to extend the transfer matrix calculation using the related density matrix renormalisation group (DMRG) method introduced by White\cite{white92,white93}, applied to two-dimensional classical spin models by Nishino\cite{nishino95} and extended to self-avoiding walk models by Foster and Pinettes\cite{FC03a,FC03b}.

The finite-size calculations rely on results from conformal invariance, which lead one naturally to calculate the scaling dimensions $x_\sigma$ and $x_\varepsilon$ with fixed boundary conditions. Translating these to the more standard exponents requires a knowledge of $\nu$. The value of $\nu$ arising from the transfer matrix calculation is at variance  with the conjectured exact value for the model. We take the opportunity of extending the original transfer matrix calculation by Bl\"ote and Nienhuis\cite{blotenienuis}. We find that, up to the lattice widths we attain, our best estimate for $x_\varepsilon=0.155$ as found by Bl\"ote and Nienhuis\cite{blotenienuis} rather than the required $x_\varepsilon =1/12=0.083333\cdots$. We conclude that either the finite-size effects are particularly severe with this particular model, or a more subtle effect is at play. Either way more work is required.

\section{Model and Transfer Matrix Calculation}

\begin{figure}
\begin{center}
\caption{A vertex interacting self-avoiding walk model showing the vertex collisions,  
weighted with a Boltzmann factor $\tau=\exp(\beta\varepsilon)$. Surface contacts are weighted $\omega_s=\exp(\beta\varepsilon_s)$ and a fugacity $K$ is introduced per walk step. The walk is shown on a strip of width $L=5$. }\label{model}

\

\includegraphics[width=10cm,clip]{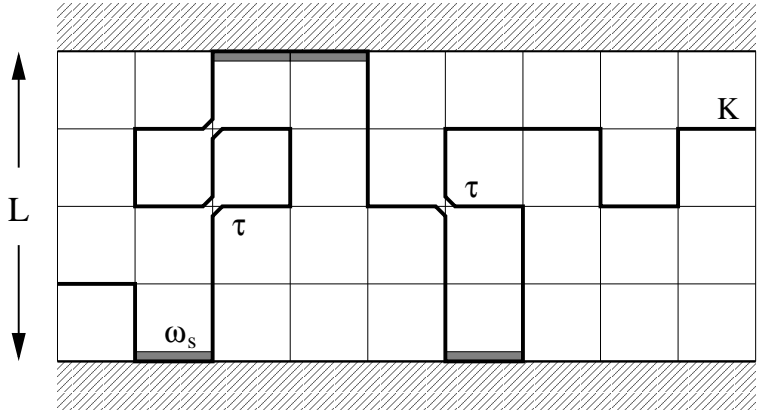}
\end{center}
\end{figure}

The model studied here is defined as follows: consider all random walks on the square lattice which do not visit any lattice bond more than once. 
The walk is not allowed to cross at doubly visited sites but 
may  ``collide". A collision is assigned an attractive energy $-\varepsilon$. 
The walk is allowed to touch, but not cross, a surface defined as a horizontal line on the lattice. 
Each step along the surface is assigned an attractive energy $-\varepsilon_S$. 
For the transfer matrix calculation that follows, it is convenient to consider a strip of width $L$ with an attractive surface both sides of the strip. 
This is not expected to change the behaviour in the thermodynamic limit $L\to \infty$; the bulk critical behaviour should not depend on the boundary conditions and when calculating the surface critical behaviour, a walk adsorbed to one surface needs an infinite excursion in order to ``see" the other surface. 
Additionally, the finite-size scaling results which link the eigenvalues of the transfer matrix to the scaling dimensions $x_\sigma^s$ and $x_\varepsilon^s$ (see \eref{sig-dim} and \eref{eng-dim}) rely on the conformal mapping of the half plane (one adsorbing surface) onto a strip with two adsorbing surfaces\cite{cardy}. A typical configuration is shown in \Fref{model}.

The partition function for the model is

\begin{equation}\label{part}
{\cal Z}=\sum_{\rm walks} K^N \omega_s^{N_S} \tau^{N_I},
\end{equation}
where $K$ is the fugacity, $\omega_s=\exp(\beta\varepsilon_S)$ and $\tau=\exp(\beta\varepsilon)$.
$N$ is the length of the walk, $N_S$ is the number of steps on the surface, and $N_I$ is the number of doubly-visited sites.

The average length of the walk is controlled by the fugacity $K$ through
\begin{equation}\label{n}
\langle N\rangle=K\frac{\partial \ln{\cal Z}}{\partial K}.
\end{equation}
As $K$ increases from zero, $\langle N \rangle$ increases, diverging at some value $K=K^\star(\omega_s,\tau)$. To start we consider what happens in the absence of the adsorbing boundary. For $\tau$  small enough, 
\begin{equation}
\langle N\rangle\sim (K^\star(\omega_s,\tau)-K)^{-1},
\end{equation}
 whilst for large enough $\tau$ the divergence is discontinuous. Whilst $\langle N\rangle$ is finite, the density of occupied bonds on an infinite lattice is zero, whilst once $\langle N \rangle$ has diverged, the density is in general finite. For small enough $\tau$ the density becomes non-zero continuously at $K=K^\star$ and for large enough $\tau$ the density jumps at $K=K^\star$. $K^\star$ may then be understood as the location of a phase transition, critical for $\tau<\tau_{\rm coll}$ and first order for $\tau>\tau_{\rm coll}$. The problem of infinite walks on the lattice is equivalent to setting $K=K^\star$ and varying $\tau$, then it may be seen that for $\tau<\tau_{\rm coll}$ the density is zero and is non-zero for $\tau>\tau_{\rm coll}$. It then follows that $\tau_{\rm coll} $ defines the collapse transition point.
 

Now let us consider the effect of the adsorbing boundary at constant $\tau$. For $\omega_s$ small, the entropic repulsion of the wall is strong enough for the walk to remain in the bulk. Once $\omega_s$ is large enough for the energy gain to overcome the entropic repulsion, the walk will visit the boundary a macroscopic number of times, and the walk adsorbs to the surface. These two behaviours are separated by $\omega_s=\omega_s^\star$. 
For $\omega_s\leq \omega_s^\star$ the behaviour of the walk is not influenced by the wall, and $K^\star$ is independent of $\omega_s$. The transition $K=K^\star$ if critical ($\tau\leq\tau_{\rm coll}$) corresponds to ordinary critical behaviour. However, for $\omega_s>\omega_s^\star$, $K^\star$ is a function of $\omega_s$, and the transition is referred to as a surface transition. The point $K=K^\star$, $\omega_s=\omega_s^\star$ is referred to as the special critical point (again $\tau\leq\tau_{\rm coll}$). 

As the critical value $K^\star$ is approached, and in the absence of a surface, the partition function \eref{part} and the correlation length $\xi$ are expected to diverge, defining the standard exponents $\gamma$ and $\nu$:
\begin{eqnarray}\label{xib}
\xi\sim|K-K^\star|^{-\nu}\\
{\cal Z}\sim|K-K^\star|^{-\gamma}
\end{eqnarray}
The effect of the surface on the walk is to introduce an entropic repulsion, pushing the walk away from the surface. The number of allowed walks is reduced exponentially if the walk is constrained to remain near the surface, in particular if one or both ends of the walk are obliged to remain in contact with the surface. In this case the divergence of $\cal Z$ is modified, and two new exponents are introduced, $\gamma_1$ and $\gamma_{11}$. Defining ${\cal Z}_1$ and  ${\cal Z}_{11}$ as the partition functions for a walk with one end, and both ends, attached to the surface respectively, then:
\begin{eqnarray}
{\cal Z}_1\sim|K-K^\star|^{-\gamma_1}\\
{\cal Z}_{11}\sim|K-K^\star|^{-\gamma_{11}}
\end{eqnarray}
Whilst the bulk exponents, such as $\nu$ and $\gamma$, are the same at an ordinary critical point and at the special critical point, the surface exponents $\gamma_1$ and $\gamma_{11}$ differ. 
The exponents $\nu$, $\gamma$, $\gamma_1$ and $\gamma_{11}$ are related by the Barber relation\cite{barber}:
\begin{equation}\label{barb}
\nu+\gamma=2\gamma_1-\gamma_{11}.
\end{equation}

The partition function may be calculated exactly on a strip of length $L_x\to\infty$ and of finite width $L$ by defining a transfer matrix ${\cal T}$. If periodic boundary conditions are assumed in the $x$-direction, the partition function for the strip is given by:
\begin{equation}
{\cal Z}_L=\lim_{L_x\to\infty}\Tr\left({\cal T}^{L_x}\right).
\end{equation}
The free energy per lattice site, the density, surface density and correlation length for the infinite strip may be calculated from the eigenvalues of the transfer matrix:
\begin{eqnarray}
f(K,\omega_s,\tau)=\frac{1}{L}\ln\left(\lambda_0\right),\\
\rho(K,\omega_s,\tau)= \frac{K}{L\lambda_0}\frac{\partial \lambda_0}{\partial K},\\
\rho_S(K,\omega_s,\tau)= \frac{\omega_s}{\lambda_0}\frac{\partial \lambda_0}{\partial \omega_s},\\\label{xi}
\xi(K,\omega_s,\tau)=\left(\ln\left|\frac{\lambda_0}{\lambda_1}\right|\right)^{-1},
\end{eqnarray}
where $\lambda_0$ and $\lambda_1$ are the largest and second largest (in modulus) eigenvalues.

Our first task is to find estimates of $K^\star(\omega_s,\tau)$. 
An estimate for the critical point where the length of the walk diverges may be found using phenomenological renormalisation group for a pair of lattice widths\cite{mpn76}, $L$ and $L^\prime$. The estimated value of $K^\star$ is given by the solution of the equation:
\begin{equation}\label{nrg}
\frac{\xi_L}{L}=\frac{\xi_{L^\prime}}{L^\prime}
\end{equation}

Both these methods give finite-size estimates $K^\star_L(\omega_s,\tau)$ which should converge to the same value in the limit $L\to\infty$. Using \Eref{xib} at the fixed point defined by \Eref{nrg}, estimates for $\nu$ and the corresponding surface correlation length exponent, $\nu_s$, 
may be calculated using 
 \begin{eqnarray}\label{nuestim}
 \frac{1}{\nu(L)}&=&\frac{\log\left(\frac{{\rm d}\xi_L}{{d}K}/\frac{{\rm d}\xi_{L+1}}{{d}K} \right)}{\log\left(L/(L+1)\right)}-1,\\\label{nusestim}
 \frac{1}{\nu_s(L)}&=&\frac{\log\left(\frac{{\rm d}\xi_L}{{d}\omega_s}/\frac{{\rm d}\xi_{L+1}}{{d}\omega_s} \right)}{\log\left(L/(L+1\right)}-1.
\end{eqnarray}

The critical dimensions of the surface magnetic and energy fields may be calculated from the first few eigenvalues of the transfer matrix: 
\begin{eqnarray}\label{sig-dim}
x^s_\sigma&=&\frac{L\ln\left|\frac{\lambda_0}{\lambda_1}\right|}{\pi},\\\label{eng-dim}
x^s_\varepsilon&=&\frac{L\ln\left|\frac{\lambda_0}{\lambda_2}\right|}{\pi},
\end{eqnarray} 
with $\lambda_2$ the eigenvalue with the third largest absolute value.

The surface scaling dimensions $x^s_\sigma$ and $x^s_\varepsilon$ may be related to the surface correlation length exponent $\nu_s$ and the exponent $\eta_\parallel$, controlling the decay of the correlation function along the surface, through standard relations
\begin{eqnarray}\label{nuref}
\nu_s&=&\frac{1}{1-x^s_\varepsilon},\\\label{eta}
\eta_{\parallel}&=& 2x^s_\sigma.
\end{eqnarray}
The entropic exponent $\gamma_{11}$ is related to $\eta_{\parallel}$ through:
\begin{equation}\label{gam11eta}
\gamma_{11}=\nu(1-\eta_\parallel).
\end{equation}

For a more detailed discussion of the transfer matrix method, and in particular how to decompose the matrix, the reader is referred to the article of Blöte and Nienhuis~\cite{blotenienuis}.

\section{Results}

The finite-size results obtained are, where possible, extrapolated on the one hand using the Burlisch and Stoer (BST) extrapolation procedure\cite{bst} and on the other hand fitting to a three point extrapolation scheme, fitting the following expression for quantity $X_L$:
\begin{equation}\label{3ext}
X_L=X_\infty+aL^{-b}.
\end{equation}
Calculating $X_\infty$, $a$ and $b$ require three lattice widths. The extrapolated values $X_\infty$ clearly will still depend weakly on $L$, and the procedure may be repeated, however weak parity effects can be seen in their values, often impeding further reasonable extrapolation by this method.

\subsection{Phase Diagram}

\begin{figure}
\begin{center}
\caption{The phase diagram calculated using the Phenomenological Renormalisation Group equation. The vertical line is placed at the best estimate of the collapse transition, expected to be independent of the surface interaction. (Colour online)}\label{pd}

\ 

\includegraphics[width=10cm,clip]{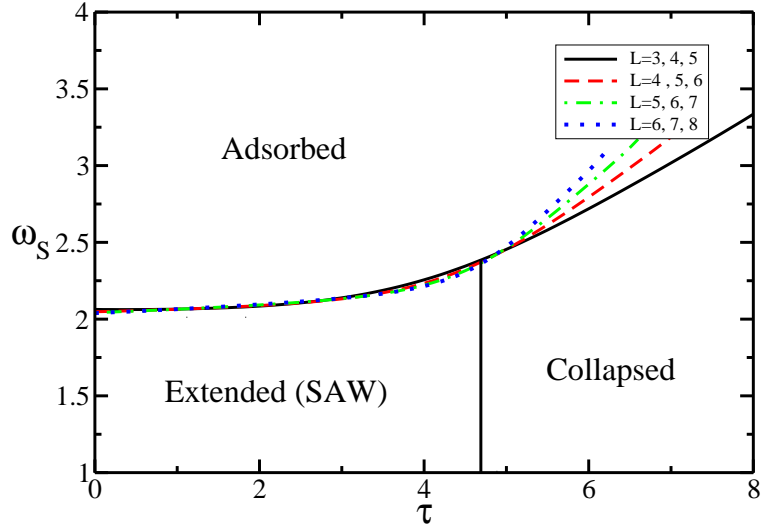}
\end{center}
\end{figure}

The phase diagram is shown as a function of $\omega_s$ and $\tau$,  projected onto the $K=K^\star(\tau,\omega_s)$ plane, in Figure~\ref{pd}. $K^\star$ is determined using equation~(\ref{nrg}) using two lattice sizes $L,L+1$. The adsorption line is then fixed by the simultaneous solution of \eref{nrg} for two sets of lattice sizes, $L,L+1$ and $L+1,L+2$, so that each line requires three lattice sizes. The vertical line is fixed from the best estimate for the bulk collapse transition, here $\tau_{\rm coll}=4.69$\cite{FC03a}.

\begin{figure}
\begin{center}
\caption{The order parameter ${\cal O}=\rho_s/(L\rho)$ plotted as a function of $\omega_s$ for $\tau=6$. (Colour online)}\label{op}

\ 

\includegraphics[width=10cm,clip]{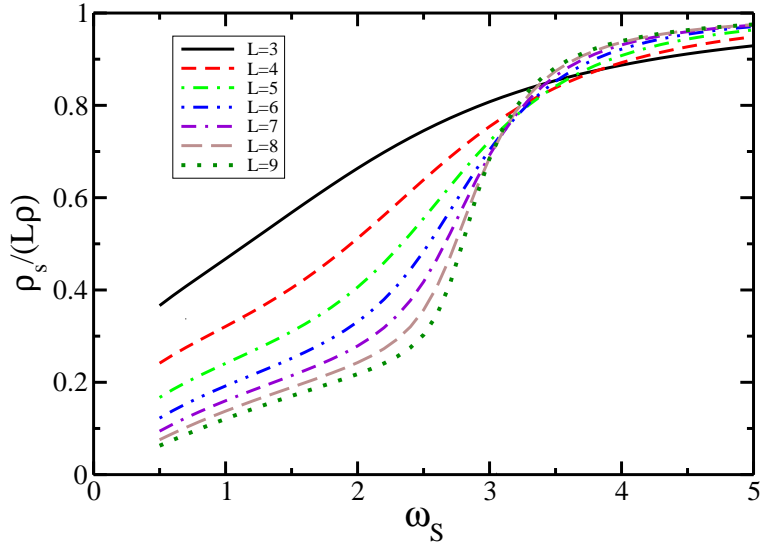}
\end{center}
\end{figure}

\begin{figure}
\begin{center}
\caption{The order parameter for a possible globule attached phase ${\cal O_I}=\rho_{i1}/(L\rho)$ is plotted as a function of $\omega_s$ for $\tau=6$. The density of interactions one row from the surface is used since it is not possible to have collisions on the surface because there are only three lattice bonds per surface site. (Colour online)}\label{rhoi}

\ 

\includegraphics[width=10cm,clip]{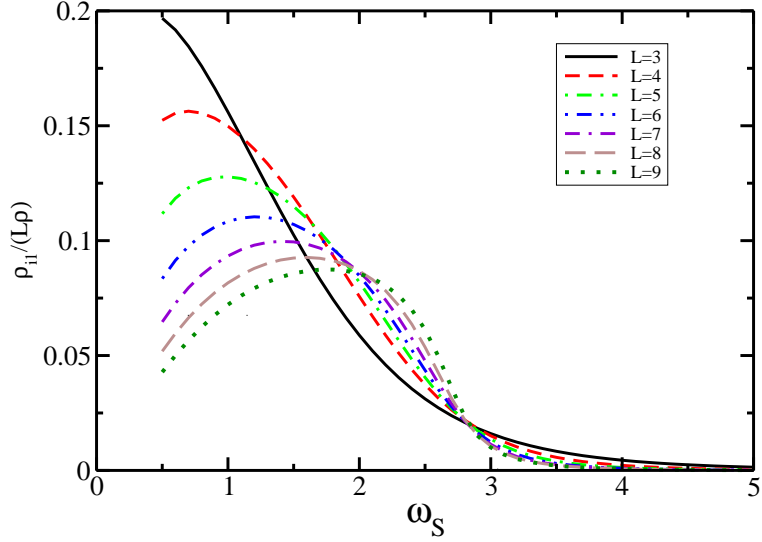}
\end{center}
\end{figure}

In the adsorbed phase, shown in the phase digram in Figure~\ref{pd}, 
the number of contacts with the surface becomes macroscopic, scaling with the length of the walk, and the density decays rapidly with the distance from the surface. 
For the $\Theta$-point model it has been shown that there is another special line in the phase diagram 
separating the collapsed phase in two: for small enough $\omega_s$ the collapsed walk avoids contacts with the wall, but for higher values of $\omega_s$ the outer perimeter  of the collapsed globule wets the surface, defining an attached globule ``phase"\cite{kumar}. 
To investigate the possibility of such a phase, we examine the order parameter for the adsorbed phase (\Fref{op}) and the density of interactions one lattice site out from the wall (there are no interactions on the wall, since four occupied bonds must collide) (\Fref{rhoi}). In the $\Theta$-point model the presence of such a phase manifests itself by a plateau in the order parameter. Such a plateau  exists, but starts at or below $\omega_s=1$, indicating that the globule is probably attached for all attractive wall interaction energies. This is consistent with the plots of the normalised density of interactions. Both plots show crossings at a value of $\omega_s$ consistent with the adsorption transition. We suggest that the entire phase is ``surface-attached", and so there is no additional line on the phase diagram shown in Figure~\ref{pd}.

\


\begin{table}
\caption{Location of the collapse value of $\tau$ with $\omega_s=1$, using three lattice widths. 
Estimates for $K_{\rm coll}$ and $\nu_{\rm coll}$ are also shown, as well as estimates for $\eta_{||}^{\rm ord}$ calculated for $\omega_s=1$ The three point extrapolations are shown in the second half of the table.}\label{ord_peq1}
\begin{indented}
\item[]\begin{tabular}{@{}lllll}
\br
$L$ & $K_{\rm coll}$ & $\tau_{\rm coll}$ & $\nu_{\rm coll}$  & $\eta^{\rm ord}_{||} $\\
\mr
3 &  0.359410 & 4.071423 & 0.614465& 1.024334 \\
4  & 0.351725 & 4.410526 & 0.596955& 1.147824\\
5  & 0.347865 & 4.540658 &  0.588407& 1.233790\\
6  &  0.345694 & 4.598914& 0.583712& 1.297254\\
7  & 0.344369 & 4.628215 & 0.580898 &  1.346184\\
8  &  0.343508 &  4.644460 & 0.579079& 1.385168\\
9  &  0.342920 & 4.654221 & 0.577827& 1.417020 \\
10 &0.342502 & 4.660572 & 0.576905& 1.443585\\
\mr
BST $\infty$ &  0.3408 &  4.673 &  0.574 & 1.77 \\
\br
\multicolumn{4}{c}{Three point extrapolated results}\\
\mr
3 &   0.339412 &4.696857 & 0.571064& 2.103158\\
4  & 0.340217 &  4.681105& 0.572693& 1.963018\\
5  & 0.340540& 4.676727 &  0.573260& 1.901623\\
6  &   0.340676& 4.676316& 0.573231& 1.868539\\
7  &  0.340749& 4.676871&  0.573019&  1.845418\\
8 & 0.340767 & 4.678731 & 0.572356& 1.832298\\
\mr
BST $\infty$ &  0.3408 &  4.69 &  0.572 & 1.77 \\
\br

\end{tabular}
\end{indented}
\end{table}

In \Tref{ord_peq1} we locate the collapse transition in strips with fixed walls and $\omega_s=1$. The collapse transition is determined  as follows: solutions to the critical line $K^\star_L(\tau)$ are found by using the phenomenological renormalisation group on a pair of lattice widths ($L$ and $L+1$) and looking for crossings in the estimates for $\nu$ for consecutive pairs of $L, L+1$. Since $\nu$ is different at the collapse point than along $\tau<\tau_{\rm coll}$ and $\tau>\tau_{\rm coll}$ lines, these estimates converge to the correct $\nu$ for the collapse transition.  This gives us the following estimates:  $K_{\rm coll}=0.3408$, $\tau_{\rm coll}=4.69$ and $\eta_{\parallel}^{\rm ord}=1.77$, as well as $\nu_{\rm coll}=0.572$.  It is noticeable that its value is much closer to that expected for the $\Theta$-point model ($\nu_{\Theta}=4/7$) than the predicted value for this model ($\nu_{O(n=0)}=12/23$). We shall see that, whilst the estimates for the other quantities of interest are remarkably stable, the estimates for $\nu_{\rm coll}$ seem rather sensitive to how they are calculated. We will return to this point later.

\begin{table}
\caption{Special point location with $\tau$ unconstrained, using four lattice widths. Estimates for $\eta_\parallel^{\rm sp}$ are also shown. The three point extrapolations are shown in the second half of the table.}\label{sp_peq1}
\begin{indented}
\item[]\begin{tabular}{@{}lllll}
\br
$L$ & $K_{\rm coll}$ & $\tau_{\rm coll}$ & $\omega_s^{\rm sp}$  & $\eta^{\rm sp}_{||} $\\
\mr
3 &  0.335871 & 4.989134 & 2.452162& -0.120915 \\
4  &  0.337720& 4.868882 & 2.418298 & -0.110883\\
5  & 0.338679 & 4.809216&  2.399537 &-0.104452\\
6  &0.339256  & 4.774456& 2.387539 & -0.099806 \\
7  &  0.339624& 4.752731 &  2.379405 &  -0.096306\\
8  &  0.339874 &4.738301  &2.373598 &-0.093563\\
9  &   0.340050& 4.728276& 2.369291&-0.091352\\
\mr
BST $\infty$ &  0.3408  & 4.6901   & 2.3513 & -0.07843\\
\br
\multicolumn{4}{c}{Three point extrapolated results}\\
\mr
3 &  0.340975  &4.682860 & 2.344211& -0.06881 \\
4  &  0.341078&  4.676046 & 2.338622 & -0.059977\\
5  & 0.340899 & 4.684292&  2.343044 &-0.064925\\
6  &  0.340848  & 4.686507&2.343975 & -0.065385 \\
7  &   0.340811& 4.688205 &  2.344840 &  -0.066177\\
\br

\end{tabular}
\end{indented}
\end{table}

In \Tref{sp_peq1} we seek the special point along the collapse transition, in other words the point at which the extended, collapsed and adsorbed phases co-exist. A different set of critical exponents is expected. In order to find the special point, we need an extra lattice width; three are required to find $\omega_s^\star(\tau)$ and a fourth is required to fix the collapse transition.
We find $K_{\rm coll}=0.3408$ and $\tau_{\rm coll}=4.69$ as in the case when $\omega_s=1$. The special point is found to be at a value of $\omega^{\rm sp}_s=2.35\pm0.01$. The estimate of $\eta_{\parallel}^{\rm sp}$ is not very precise, but here seems to be around $-0.06 \to -0.07$.

\subsection{Results for the semi-flexible VISAW}

The model may be extended by introducing different weighting for the corners and the straight sections. 
We follow the definitions in Reference\cite{blotenienuis} and add a weight $p$ for each site where the walk does not take a corner (i.e. for the straight sections). 
As $p$ is varied we expect the collapse transition point to extend into a line. It turns out that there is an exactly known point along this line.
The location of this point is given
exactly as\cite{blotenienuis}:
\begin{equation}\label{exactpt}
\left.\begin{array}{rcl}
z=K^2\tau&=&\left\{2-\left[1-2\sin(\theta/2)\right]
\left[1+2\sin(\theta/2)\right]^2\right\}^{-1}\\
K&=&-4z\sin(\theta/2)\cos(\pi/4-\theta/4)\\
pK&=&z\left[1+2\sin(\theta/2)\right]\\
\theta&=&-\pi/4
\end{array}\right\}.
\end{equation}
This gives exactly the location of the multicritical collapse point when $p=p^\star=0.275899\cdots$ as
$K_{\rm coll}=0.446933\cdots$ and
$\tau_{\rm coll}=2.630986\cdots$.  
Using this exactly known point we hope to be able to extend the number of different data points and improve the precision of the determination for different surface exponents.

In \Tref{ord_pstar} we calculate estimates for $\eta_\parallel^{\rm ord}$ in two ways. Firstly we fix $\omega_s=1$ and $p=p^\star$ and determine $K^\star$ by solving \Eref{nrg} and determining the multicritical point by looking for crossings in the estimates for $\nu$. Fixing the multicritical point this way requires three lattice sizes, $L, L+1$ and $L+2$. Estimates for $K_{\rm coll}$ and $\tau_{\rm coll}$ calculated in this way are shown in the columns marked {\bf A} and are seen to converge nicely to the expected values. The second method used consisted in fixing $K$, $\tau$ and $p$ to their exactly known multicritical values, and fixing $\omega_s$ to the ordinary fixed point looking for solutions to \Eref{nrg}. This only requires two lattice widths, giving an extra lattice size. The values of $\eta_\parallel^{ord}$ are shown as calculated from the two methods, and converge to values consistent with the $p=1$ case. 

\begin{table}
\caption{Table shows:\\
 {\bf A} the location of the value of $\tau_{\rm coll}$ with $\omega_s=1$ and $p=p^\star$, 
 using three lattice widths and crossings of  $\nu$ estimates. Estimates for $K_{\rm coll}$ and $\eta_\parallel^{\rm ord}$ are also shown.\\
{\bf B} Estimates for $\omega_s$ for the ordinary point along with the estimates for $\eta_\parallel^{\rm ord}$ at this point with $K=K_{\rm coll}$, $p=p^\star$ and $\tau=\tau_{\rm coll}$. 
Only two lattice sizes are required to fix $\omega_s$, hence we can calculate up to a lattice width of $L=11$.}\label{ord_pstar}
\begin{indented}
\item[]\begin{tabular}{@{}l|lll|lll}
\br
 & \multicolumn{3}{c|}{A} & \multicolumn{2}{c}{B}\\\mr
$L$ & $K_{\rm coll}$ & $\tau_{\rm coll}$   & $\eta^{\rm ord}_{||} $ & $\omega^{\rm ord}_s$ & $\eta_\parallel^{\rm ord}$\\
\mr
3 &0.464018  &2.309912 & 1.401892&  0.760808 &1.813498\\
4  & 0.457207& 2.451700 &1.471983&0.785333 &1.787052\\
5  & 0.453616& 2.520291& 1.520159&0.797646 &1.776227\\
6  &  0.451604 &2.556015 &1.554337& 0.805442 &1.770439 \\
7  & 0.450391 &2.576330  &1.579631  & 0.811096 & 1.766807\\
8  &0.449615  & 2.588789 &1.599034& 0.815550 &1.764286\\
9  & 0.449089 & 2.596962 & 1.614375& 0.819245 & 1.762416 \\
10 &0.448720 & 2.602583  &1.626774 & 0.822417 & 1.760965\\
11 & --- & --- & --- & 0.825204 & 1.759802 \\
\mr
BST $\infty$ & 0.4473 & 2.597 & 1.708 & 0.8955 &1.7499\\
exact  &  $ 0.446933\cdots$ & $2.630986\cdots$  & \\
\br
\multicolumn{6}{c}{Three point extrapolated results}\\
\mr
3 & 0.444582 &2.656251 & 1.953742& 0.824571& 1.761506\\
4 & 0.446572 &2.628298 & 1.807661& 0.789157& 1.757867\\
5 &  0.447052& 2.622897& 1.774328& 0.799994& 1.755193\\
6 & 0.447197 & 2.621984& 1.762266& 0.863901&1.753208 \\
7 & 0.447252 &2.622163 & 1.754657& 0.881961&1.751795 \\
8 & 0.447224 & 2.623411&1.753850 & 0.816627& 1.750839\\
9 & --- & --- & --- & 0.820168& 1.750223\\
\br

\end{tabular}
\end{indented}
\end{table}

In \Tref{etas_pstar} we present results calculated fixing $\tau=\tau_{\rm coll}$ and looking for simultaneous solutions of the phenomenological renormalisation group equation \eref{nrg}. These solutions exist at two values of $\omega_s$, the ordinary and the special fixed points. The values of $K_{\rm coll}$, $\omega_s$ 
and $\eta_\parallel$ are given for the two fixed points. Again agreement is found with previous values calculated.

\begin{table}
\caption{Location and estimates of $\eta_\parallel$ for both the ordinary and special point  fixing $\tau=\tau_{\rm coll}(p^\star)$, and using three lattice widths. 
The three point extrapolations are shown in the second half of the table.}\label{etas_pstar}
\begin{indented}
\item[]\begin{tabular}{@{}l|lll|lll}
\br
 & \multicolumn{3}{c|}{Ordinary Point} & \multicolumn{3}{c}{Special Point}\\\mr
$L$ & $K_{\rm coll}$  & $\omega^{\rm ord}_s$  & $\eta_{||}^{\rm ord}$ & $K_{\rm coll}$  & $\omega^{\rm sp}_s$  & $\eta_{||}^{\rm sp}$ \\
\mr
3 & 0.444849 &0.727730& 1.876811 & 0.444289  & 3.840487& -0.254004 \\
4  & 0.446261& 0.765809& 1.816907 &0.445726 & 3.660264 & -0.191557 \\
5  &0.446626&0.782852 & 1.795039  & 0.446279&3.575039 &  -0.157742 \\
6  & 0.446763 &0.792806&  1.784181&  0.446537 &3.527515&   -0.136796 \\
7  &0.446826 &0.799587&1.777736 & 0.446675 & 3.498075&  -0.122630 \\
8  &  0.446861&0.804688  & 1.773440 &  0.446754 & 3.478454&  -0.112441\\
9  & 0.446881&0.808784& 1.770341 &  0.446804& 3.464654& -0.104773\\
10 & 0.446894 &0.812223& 1.767979& 0.446836 & 3.454537& -0.098797\\
\mr
BST $\infty$ &  0.446933  &0.8529   & 1.75 &0.44693 &3.4029 &-0.05241 \\
exact & $0.446933\cdots$& & & & & \\
\br
\multicolumn{7}{c}{Three point extrapolated results}\\
\mr
3 & 0.445398&0.740647& 1.855322& 0.444797 &3.415470& -0.065260\\
4  &0.446899&0.821222& 1.809568&0.445915 & 3.410716& -0.062347\\
5  &0.446915& 0.830707& 1.791541 &0.446934&3.506065 & -0.060003 \\
6  & 0.446923&0.794846& 1.782160& 0.446933&3.407005& -0.058240  \\
7  & 0.446928 &0.801100 & 1.755376&0.446932&3.406126&  -0.056866\\
8  &0.446931 &  0.867939 & 1.753498 &0.446932&3.405498&  -0.055762\\
\br

\end{tabular}
\end{indented}
\end{table}



\begin{table}
\caption{Location of the special point for $K=K_{\rm coll}$, $p=p^\star$ and $\tau=\tau_{\rm coll}$ 
using intersections of 
 $\eta_{||}^{\rm sp}$. The three point extrapolations are shown in the second half of the table.
 The values of $x_{\varepsilon}^s$ are calculated in the even sector at the value of $\omega^{\rm sp}_s$.}\label{nu_sp_s}
\begin{indented}
\item[]\begin{tabular}{@{}lllllllll}
\br
$L$ & $\omega^{\rm sp}_s$ & $\eta_{||}^{\rm sp}$ & $\nu$  & $\nu^{\rm sp}_{s} $ & $\phi_s=\nu/\nu_s$ & $x_\varepsilon^s(L)$ &  $x_\varepsilon^s(L+1)$ \\
\mr
3 & 3.575571 & -0.170489&0.527116& 1.829877 & 0.288061 & 0.527910&0.422737\\
4  & 3.506382& -0.135551&0.534983& 1.725710& 0.310008  & 0.449880&0.401500\\
5  & 3.472677&-0.116031&0.540007&1.668509 & 0.323646 & 0.416568&0.388462\\
6  & 3.453634& -0.103697&0.543502& 1.632771&0.332874 & 0.397954&0.379518\\
7  &3.441756 &-0.095233&0.546069&1.608442 & 0.339497&0.386019 &0.372974\\
8  &3.433803 & -0.089075&0.548029&1.590960 & 0.344477 & 0.377697&0.367970\\
9  &3.428190 &-0.084396&0.549571& 1.577675& 0.348349& 0.371554&0.364018\\
10 & 3.424062 &-0.080719&0.550811&1.567260 & 0.351393& 0.366832&0.360818\\
11 &3.420926 &-0.077751&0.551828 & 1.558893&  0.353944 &0.363087 &0.358175\\
\mr
BST $\infty$ &  3.402 & -0.0499 & 0.5592  & 1.487& 0.3777  & 0.332 & 0.332 \\
\br
\multicolumn{4}{c}{Three point extrapolated results}\\
\mr
3 &   3.404662& -0.056758&0.567274& 1.504747& & 0.367160& 0.330606\\
4  &3.404891 &-0.056025 &0.566011&1.501149& & 0.351913& 0.326878 \\
5  &3.404606&-0.054962 & 0.564760&1.497315& & 0.344469&0.326617 \\
6  & 3.404268 &-0.053981&0.563832&1.498468& &0.340278 & 0.327150 \\
7  & 3.403967&-0.053136& 0.563148& 1.488934& &0.337719 & 0.327793 \\
8 & 3.403715& -0.052418&0.562652&1.488636&  & 0.336068& 0.328375\\
9 & 3.403508 & -0.051807 &0.562224&1.488766&  & 0.334959& 0.328865 \\
\br

\end{tabular}
\end{indented}
\end{table}



At the ordinary point the exponent $\nu_s$ is expected trivially to take the value $-1$, and this was verified in the various calculations at the ordinary point, with good convergence. 
At the special point the correlation length along the surface is not expected to be trivial. In order to obtain the best estimate for this exponent we determined the location of the special point by fixing $K$, $\tau$ and $p$ to their multicritical values and then determining the special point by looking again for solutions to the phenomenological renormalisation equation \eref{nrg}, shown in \Tref{nu_sp_s} and using \Eref{nuestim} and \Eref{nusestim}.
One may also obtain an independent estimate for $\nu_s$ calculating the scaling dimension $x_\varepsilon^s$ using \eref{eng-dim}, from which $\nu_s=(1-x_\varepsilon^s)^{-1}$. The special point was determined using the odd sector of the transfer matrix, whereas the calculation of $x_\varepsilon^s$ requires the even sector, so whilst the determination method only gives one estimate of $\eta_\parallel^{\rm sp}$ it gives two estimates (one for each lattice size) for the critical dimension $x_\varepsilon^s$. These different estimates are shown in \Tref{nu_sp_s}. The values of $\nu^{\rm sp}_s$ converge to $1.487$ whilst $x_\varepsilon=0.332$, which gives $\nu^{\rm sp}_s=1.497$. It seems likely that $\nu^{\rm sp}_s=1.49\pm0.01$. Again the estimates of $\nu$ do not converge to $\nu=12/23$, but neither do they converge to the values found 
above for $\omega_s=1$ and $p=1$ (see \Tref{ord_peq1}). The crossover exponent is calculated from the estimates found for each size $\phi_s=\nu/\nu_s$, therefore the extrapolated value is only as good as the estimated values of $\nu$ and $\nu_s$. If we believe $\nu=12/23$ and $\nu_s=1.5$ then $\phi_s=8/23=0.34782\cdots$.

\section{Extending the results with DMRG}

One of the limitations of the transfer matrix method is the limited number of lattice widths that may be investigated. One way of getting round this problem is to generate approximate transfer matrices for larger widths. There exists a method of choice for doing this; 
the Density Matrix Renormalisation Group Method (DMRG) introduced by White\cite{white92,white93}, extended to classical 2d models by Nishino\cite{nishino95} and self-avoiding walk models by Foster and Pinettes\cite{FC03a,FC03b}.

The DMRG method constructs approximate transfer matrices for size $L$ from a transfer matrix approximation for a lattice of size $L-2$ by adding two rows in the middle of the system. This process is clearly local, whereas the VISAW walk configurations are clearly non-local. This problem is solved by looking at the model as the limit $n\to 0$ of the $O(n)$ model.

\begin{figure}
\begin{center}
\caption{Local complex vertices for the DMRG method}

\ \\

\includegraphics[width=10cm]{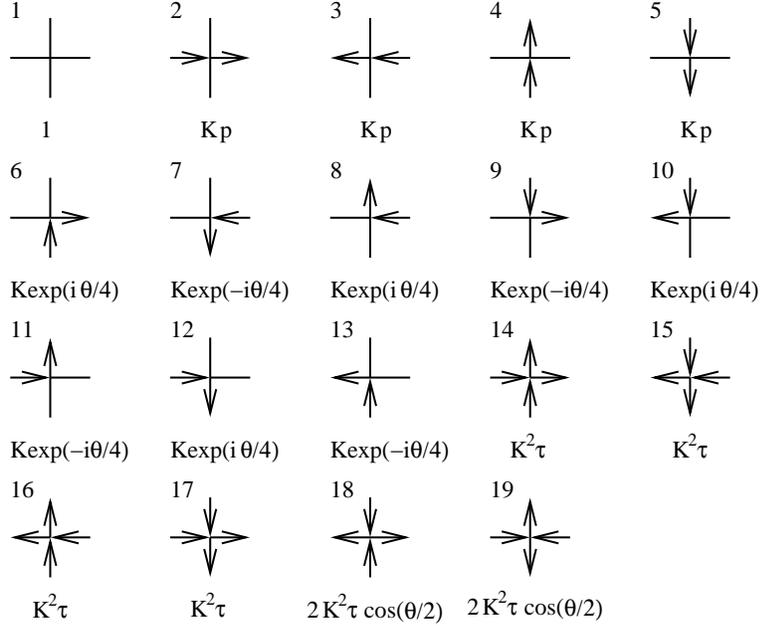}\label{vertices}
\end{center}
\end{figure}

The partition function of the $O(n)$ model is given by:
\begin{equation}
{\cal Z}_{O(n)}=\sum_{\cal G} n^{l} K^N p^{N_{\rm st}} \omega_s^{N_S}\tau^{N_I},
\end{equation}
where ${\cal G}$ denotes the sum over all graphs containing loops which visit lattice bonds at most once and which do not cross at lattice sites and $l$ is the number of such loops. $N_{\rm st}$ is the number of straight sections. In the limit $n\to 0$ the model maps onto the expected model with the odd sector of the corresponding transfer matrix giving the walk graphs, as above. Viewing the model in this way enables us to map the loop graphs into oriented loop graphs. Each loop graph corresponds to $2^{N_{\rm loops}}$ oriented loop graphs.  We associate different weights $n_+$ and $n_-$ for the different orientations, with $n=n_++n_-$, and this enables us to rewrite the partition function as follows:
\begin{eqnarray}
{\cal Z}_{O(n)}&=&\sum_{\cal G} \left(n_++n_-\right)^{l} K^N p^{N_{\rm st}} \omega_s^{N_S}\tau^{N_I}\\
&=&\sum_{\cal G^\star} n_+^{l_+}n_-^{l_-} K^N p^{N_{\rm st}} \omega_s^{N_S}\tau^{N_I}.
\end{eqnarray}
Whilst the weights $n_+$ and $n_-$  are still not local, they may be made local by realising that
four more corners in one direction than the other are required to close a loop on the square lattice. If we associate $\alpha$ with each clockwise corner and $\alpha^{-1}$ for each anti-clockwise corner we find $n_+=\alpha^4$ and $n_-=\alpha^{-4}$, setting $\alpha=\exp(i\theta/4)$ gives:
\begin{equation}
n=\alpha^4+\alpha^{-4}=2\cos\theta.
\end{equation}
The model studied here then corresponds to $\theta=\pi/2$. The resulting local (complex) weights are shown in \Fref{vertices}.

\begin{figure}
\begin{center}
\caption{Figure shows the schematic transfer matrix obtained from DMRG iteration. Circles show spins defined for the original model (3 state spins for the lattice bond: empty, and two arrow states). Squares show the m-state pseudo spins. The open circles are summed. The projection of the upper half is also shown schematically.  }\label{DMRG}
\ \\
\includegraphics[width=10cm,clip]{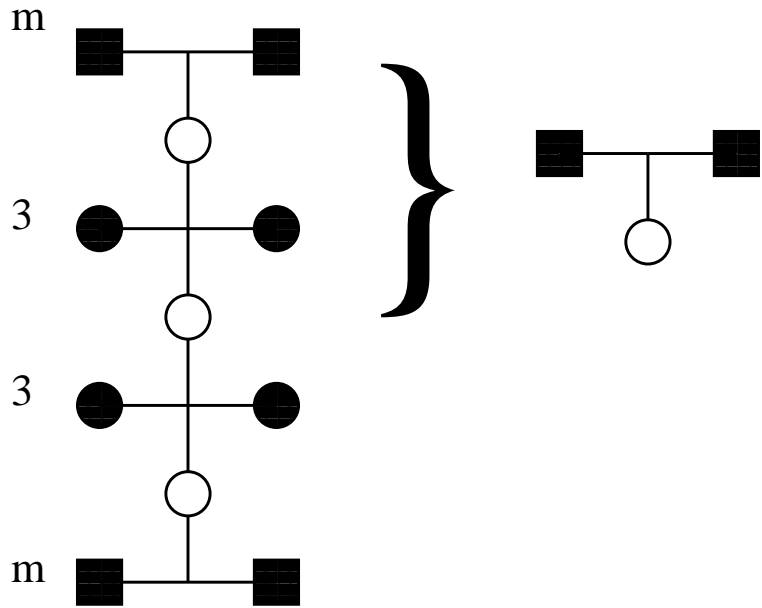}

\end{center}
\end{figure}

Now that the vertices are local, the DMRG method may be applied. The vertices represented in \Fref{vertices} may be most easily encoded by defining a three state spin on the lattice bonds, for the horizontal bonds the three states would be arrow left, empty and arrow right.
For details of the DMRG method the reader is referred to the articles by White\cite{white92,white93} and  Nishino\cite{nishino95}, but in essence the method consists in representing the transfer 
matrix for the VISAW model as the transfer matrix for an equivalent system where the top and bottom of the strip is represented by an $m$-state pseudo-spin with only the inner two of rows 
kept explicitly in terms of the original 3 state spins. For small lattice widths this identification may be done exactly, and the interaction matrix may be chosen exactly, but for a fixed value of $m$ there will come a stage where this procedure is no longer exact. At this stage the phase space in the $m$-spin representation is smaller than for the real system and an approximation must be made. Starting from the largest lattice width that may be treated exactly by the pseudo-spin system, two vertices are inserted in the middle (see \Fref{DMRG}). The $3\times m$ states at the top of the system must be projected onto $m$ states to recover a new pseudo-spin system. This must be done so as to lose the smallest amount of information, and this is where the DMRG method comes in. It turns out that the best change of basis  is derived by constructing the density matrix for the top half of the lattice strip from the ground-state eigenvector by tracing over the lower half system. The density matrix is then diagonalised and the $m$ basis vectors corresponding to the $m$ largest eigenvalues of the density matrix are kept. 

\begin{figure}
\begin{center}
\caption{Location of the ordinary collapse point (A) and corresponding value of $\eta^{\rm ord}_{\parallel}$ (B) with $p=p^\star$, $K=K_{\rm coll}$ and $\tau=\tau_{\rm coll}$. (Colour online)}
\label{DMRGord}
\ \\
{\bf A} 
\ \\ 
\includegraphics[width=10cm,clip]{wsord}
\ \\
{\bf B} 
\ \\ 
\includegraphics[width=10cm,clip]{etaord}
\end{center}
\end{figure}

\begin{figure}
\begin{center}
\caption{Location of the special collapse point (A) and corresponding value of $\eta^{\rm sp}_{\parallel}$ (B) with $p=p^\star$, $K=K_{\rm coll}$ and $\tau=\tau_{\rm coll}$. (Colour online)}\label{DMRGsp}
\ \\
{\bf A} 
\ \\ 
\includegraphics[width=10cm,clip]{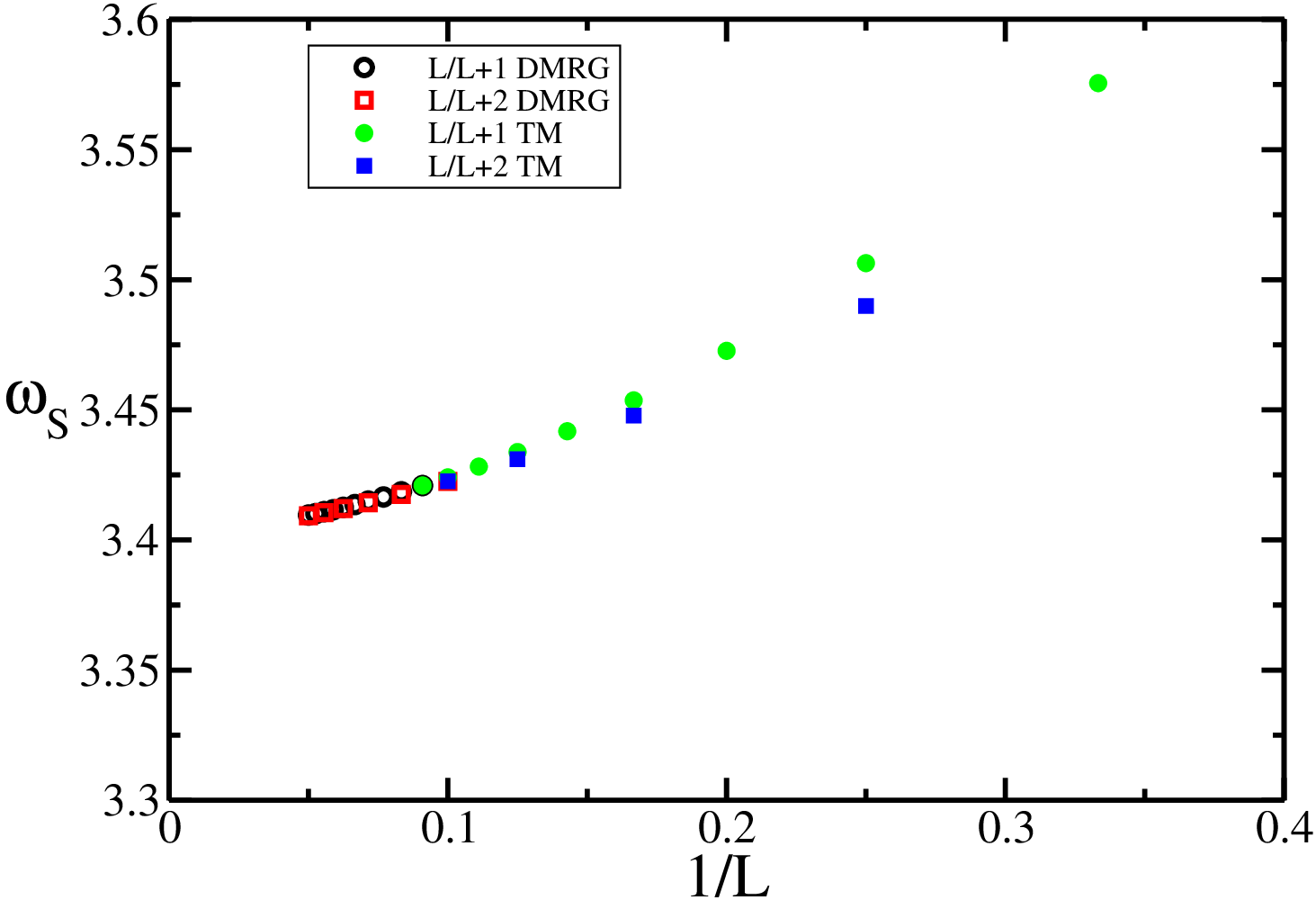}
\ \\
{\bf B} 
\ \\ 
\includegraphics[width=10cm,clip]{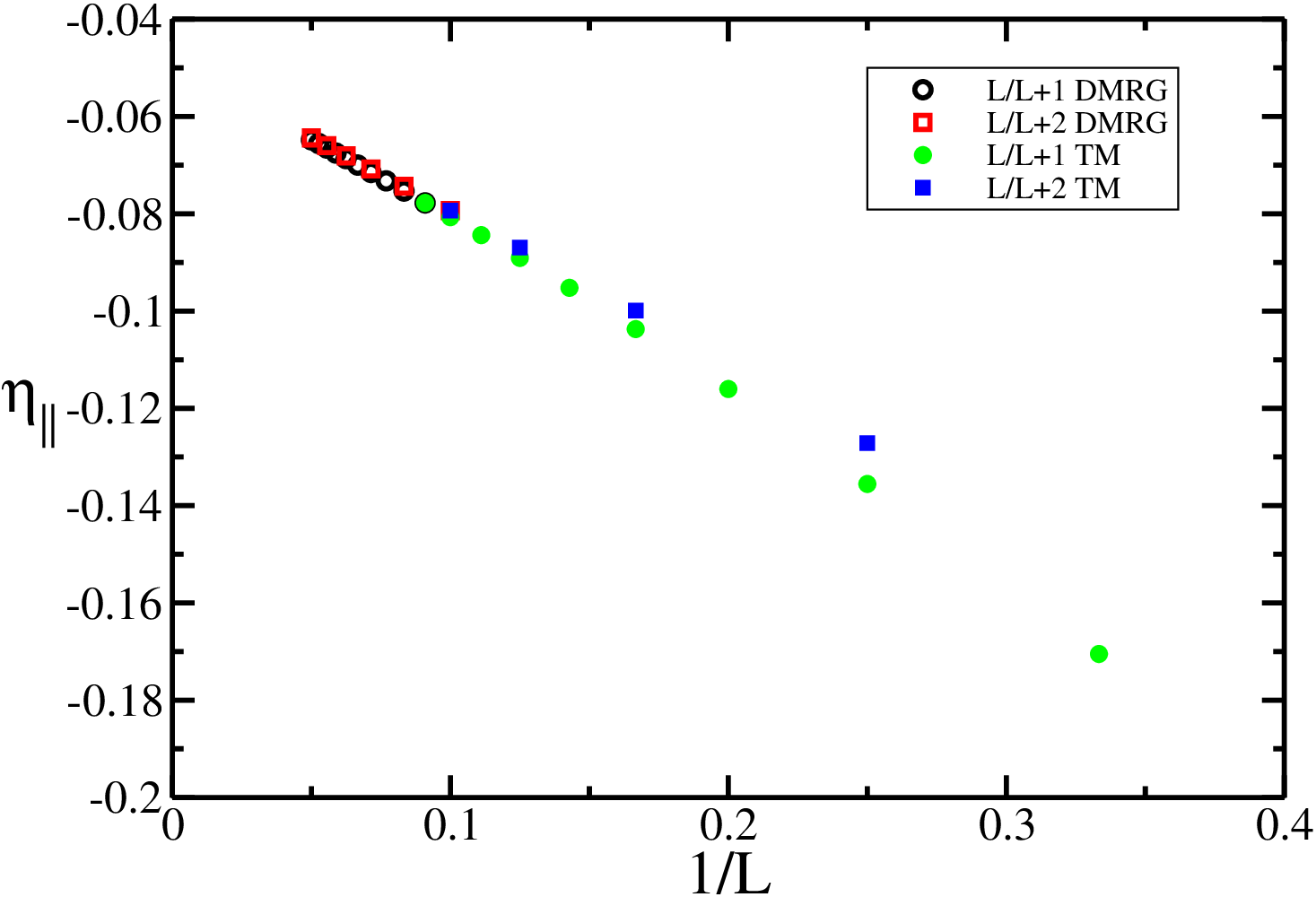}
\end{center}
\end{figure}

There are two modifications on the basic method which improve the quality of the results obtained. The number of left arrow minus the number of right arrows is conserved from one column to the next, so the transfer matrix may be split into sectors which are much smaller than the original. Since the DMRG method may be viewed as a variational method, the quality of the results may be improved by using the scanning (or finite size method) DMRG method where once the desired lattice width is obtained one grows one half of the system and shrinks the other, whilst projecting as before, so that the exactly treated spins move across the system. As few as three or four sweeps is known to vastly improve the precision of the method\cite{white92,white93}.

\begin{figure}
\begin{center}
\caption{Calculation of $x_\varepsilon^s$ using DMRG at the special collapse transition with $p=p^\star, K=K_{\rm coll}$ and $\tau=\tau_{\rm coll}$. (Colour online)}\label{DMRGxepsi}
\includegraphics[width=10cm,clip]{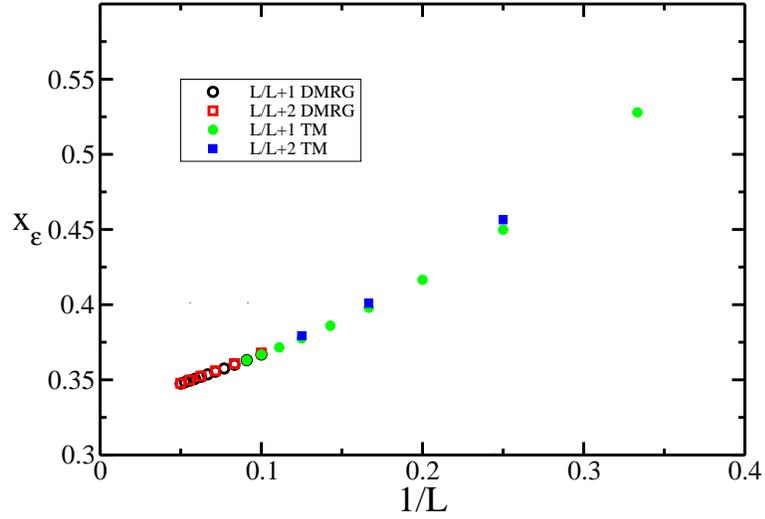}
\end{center}
\end{figure}

Clearly the precision of the method is controlled by $m$; the larger $m$ the greater the information kept. In what follows we varied $m$ up to values of $m=200$ and verified that good convergence was obtained. This conditioned the lattice size we looked at in DMRG. Whilst physical quantities such as the density converge rapidly with $m$, scaling dimensions (which interest us here) converge more slowly. As a result the largest lattice width presented here is $L=20$, which nevertheless corresponds to a good improvement over the pure transfer matrix method. 

For the DMRG calculation we fixed $p=p^\star$; $K=K_{\rm coll}$; $\tau=\tau_{\rm coll}$ 
and used the solutions of \Eref{nrg} to find the ordinary and special fixed points as well as the corresponding $\eta_\parallel$. The $x_\varepsilon^s$ were calculated from the even sector at these fixed points.

In \Fref{DMRGord} and \Fref{DMRGsp} we show the DMRG results along with the transfer matrix results for $\omega_S$ and $\eta_\parallel$ for the ordinary and special points. We deduce for the ordinary point $\omega_S^{\rm ord}=0.86\pm0.01$ and $\eta_\parallel^{\rm ord}=1.75\pm 0.01$ and for the special point $\omega_S^{\rm sp}=3.41\pm0.01$ and $\eta_\parallel^{\rm sp}=-0.05\pm0.01$.

In \Fref{DMRGxepsi} we show the estimates for the scaling dimension $x_\varepsilon^s$ calculated at the special point. We determine $x_\varepsilon^s=0.333\pm0.001$. This leads to $\nu^{\rm sp}_s=1.5$.

\section{Discussion}

\begin{figure}
\begin{center}
\caption{Bulk scaling dimensions $x_\sigma$ and $x_\varepsilon$ calculated at $p=p^\star$, $K=K_{\rm coll}$ and $\tau=\tau_{\rm coll}$ for periodic boundary conditions using DMRG with $m$ up to $m=190$. (Colour online)}\label{DMRGperiodic}

\ \\

\includegraphics[width=10cm,clip]{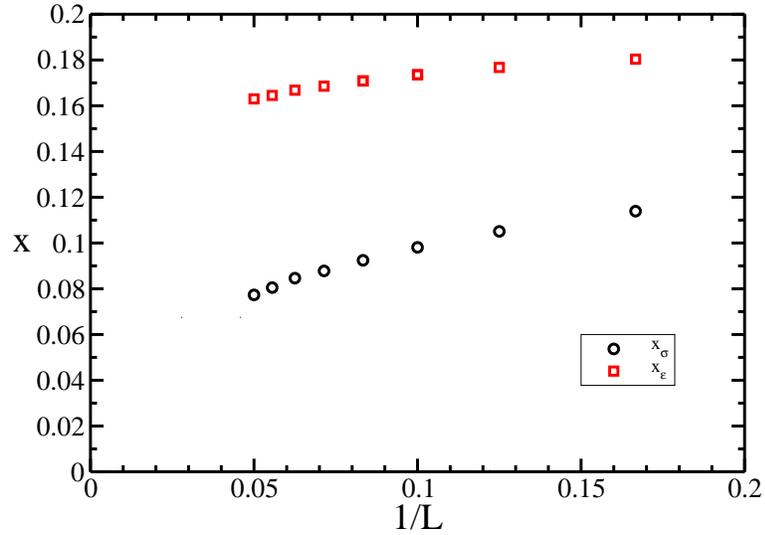}
\end{center}
\end{figure}

To conclude, we summarise the exponent values found:
\begin{center}
\begin{tabular}{ccccc}
\br
Method & $\eta_\parallel^{\rm ord}$ & $\eta_\parallel^{\rm sp}$ & $\nu_s^{\rm sp}$ & $x_\varepsilon^s$\\
\mr
TM& $1.75\pm 0.05$ & $-0.05 \to -0.08$ & $1.48\pm0.04$ & $0.332\pm 0.005$\\
DMRG&$1.75\pm 0.01$ & $-0.05 \pm 0.01$ &---& $ 0.333\pm 0.001$\\
\mr
\end{tabular}
\end{center}
As discussed above, the calculation of the exponents $\gamma_1$ and $\gamma_{11}$ as well as $\phi_s$ require a knowledge of the bulk  exponents $\nu$ and $\gamma$. Whilst there are 
conjectured exact values for these exponents, and in particular the exponent $\nu=12/23$ is found to a good level of precision for the Trails model which tends to lend support to this value, the transfer matrix calculations for the VISAW walk model do not seem to reproduce the required values, Bl\"ote and Nienhuis\cite{blotenienuis} find $x_\varepsilon=0.155$ rather than the required $x_\varepsilon=1/12$ for example.
We extend the transfer matrix results for the periodic boundary conditions using DMRG, and find a result for $x_\varepsilon$ consistent with Bl\"ote and Nienhuis\cite{blotenienuis}(see \Fref{DMRGperiodic}).

Further work is required to calculate the exponents by different methods, for example Monte Carlo, in order to understand the apparent differences in results which arise in the exponent $\nu$. Either the differences are a result of particularly strong finite-size scaling effects, which is surprising since the surface exponents themselves seem to be remarkably stable in comparison, or perhaps an indication that the critical behaviour of this model is more subtle than initially thought. Either way the model warrants further study.


\


\begin{thebibliography}{12}
\bibitem{flory} P. Flory \emph{Principles of Polymer Chemistry}, Ithaca: Cornell University Press, 1971
\bibitem{vanderbook} C. Vanderzande, \emph{Lattice Models of Polymers}, Cambridge: CUP, 1998
\bibitem{degennes75} P.~G. de Gennes, \emph{J. Phys. Lett} {\bf 36} L55 (1975)
\bibitem{ds} B. Duplantier and H. Saleur, \emph{Phys. Rev. Lett.} {\bf 59} 539, (1987)
\bibitem{seno88} F. Seno, A. L. Stella and C. Vanderzande, \emph{Phys. Rev. Lett.} {\bf 61} 1520 (1988)
\bibitem{dsb} B. Duplantier and H. Saleur, \emph{Phys. Rev. Lett.} {\bf 61} 1521, (1988)
\bibitem{merovitchlima} H. Meirovitch and H.~A. Lim, \emph{Phys. Rev. Lett}, {\bf 62} 2640 (1989)
\bibitem{dsc} B. Duplantier and H. Saleur, \emph{Phys. Rev. Lett.} {\bf 62} 2641, (1989)
\bibitem{vss91} C. Vanderzande, A. L Stella and F. Seno, \emph{Phys Rev Lett} {\bf 67} 2757 (1991) 
\bibitem{foster92} D. P. Foster, E. Orlandini and M. C. Tesi \emph{J. Phys A} {\bf 25} L1211 (1992)
 \bibitem{veal} A. R. Veal, J. M. Yeomans and G. Jug \emph{J. Phys. A} {\bf 24} 827 (1991)
\bibitem{ss88} F. Seno and A. L Stella, \emph{Europhysics Lett} {\bf 7} 605 (1988)
\bibitem{blotenienuis} H.~W.~J. {Bl\"ote} and B.~Nienhuis, \emph{J. Phys. A}, {\bf 22}  1415, (1989)
\bibitem{wbn} S.~O. Warnaar, M.~T. Batchelor, and B.~Nienhuis, \emph{J. Phys. A}, {\bf 25} 3077, (1992)
\bibitem{massih75} A.~R. Massih and M. A. Moore, \emph{J. Phys. A}, {\bf 8} 237, (1975)
\bibitem{F09} D.~P. Foster,  \emph{J. Phys. A}, {\bf 42} 372002 (2009)
\bibitem{bradley89} R. M. Bradley \emph{Phys Rev A} {\bf 39} R3738 (1989)
\bibitem{bradley90} R. M. Bradley \emph{Phys Rev A} {\bf 41} 914 (1990)
\bibitem{F10} D.~P. Foster,  \emph{J. Phys. A}, {\bf 43} 335004 (2010)
\bibitem{white92} S.R. White, \emph{Phys. Rev. Lett.} {\bf 69} 2863 (1992)
\bibitem{white93} S.R. White, \emph{Phys. Rev. B} {\bf 48} 10345 (1993)
\bibitem{nishino95} T. Nishino, \emph{J. Phys. Soc. Jpn.} {\bf 64} 3598 (1995)
\bibitem{FC03a} D.~P. Foster and C. Pinettes \emph{J. Phys. A}, {\bf 36} 10279 (2003)
\bibitem{FC03b} D.~P. Foster and C. Pinettes \emph{Phys Rev E}, {\bf 67} R045105 (2003)
\bibitem{cardy} J. Cardy in \emph{Phase Transitions and Critical Phenomena}, eds. Domb and Lebowitz, {\bf Vol. XI} (Academic, New York), 1986\bibitem{barber} M.~N. Barber, \emph{Phys. Rev B} {\bf 8} 407 (1973)
\bibitem{derh} B. Derrida  and H. G. Herrmann  \JP {\bf 44} 1365 (1983)
\bibitem{mpn76} M. P. Nightingale  {\it Physica} {\bf A83} 561(1976)
\bibitem{bst} R. Bulirsch and J. Stoer, \emph{Numer. Math.} {\bf 6} 413 (1964);
 M. Henkel and G. Schütz \emph{J. Phys. A} {\bf 21} 2617 (1988)
\bibitem{kumar} Y. Singh, D. Giri and S. Kumar \emph{J. Phys A} {\bf 34} L67 (2001) 
\end{thebibliography}

\end{document}